\documentclass[12pt]{iopart}
\usepackage{amssymb}
\usepackage{textcomp}
\usepackage[dvips]{graphicx}
\usepackage[numbers,sort&compress]{natbib}
\begin{document}

\title{High-field electron transport in bulk ZnO}

\author{L.~Ardaravi\v{c}ius$^1$\footnote[1]{To whom correspondence should be addressed
(linas.ardaravicius@ftmc.lt)}, O.~Kiprijanovi\v{c}$^1$,
J.~Liberis$^1$, M.~Ramonas$^1$, E. \v{S}ermuk\v{s}nis, A.
Matulionis$^1$, M. Toporkov$^{2}$, V.~Avrutin$^{2}$,
\"U.~\"Ozg\"ur$^{2}$ and H.~Morko\c{c}$^{2}$}

\address{$^1$Center for Physical Sciences and Technology,
Saul\.{e}tekio 3, Vilnius, Lithuania}
\address{$^2$Department of Electrical and Computer Engineering, Virginia Commonwealth University, Richmond,
Virginia 23284, USA}

\date{\today}% It is always \today, today,
\begin{abstract}

Current-voltage dependence is measured in (Ga,Sb)-doped ZnO up to
150 kV/cm electric fields. A channel temperature is controlled by
applying relatively short (few ns) voltage pulses to two-terminal
samples. The dependence of electron drift velocity on electron
density ranging from 1.42$\times$10$^{17}$ cm$^{-3}$ to
1.3$\times$10$^{20}$ cm$^{-3}$ at a given electric field is deduced
after estimation of the sample contact resistance and the Hall
electron mobility. Manifestation of the highest electron drift
velocity up to $\sim$1.5$\times$10$^{7}$ cm/s is estimated for
electron density of 1.42$\times$10$^{17}$ cm$^{-3}$ and is in
agreement with Monte Carlo simulation when hot-phonon lifetime is
below 1 ps. A local drift velocity maximum is observed at
$\sim$1.1$\times$10$^{19}$ cm$^{-3}$ and is in agreement with
ultra-fast hot phonon decay.

\end{abstract}

\maketitle

\section{Introduction}

Semiconducting zinc oxide (ZnO) seems to be a potential material for
optical and electronic applications \cite{morkoc2009}. It has a
relatively wide bandgap (3.37 eV) similar to GaN and is transparent.
Expected high structural quality promises efficient operation of
high power ZnO field-effect transistors at centimeter and
millimeter-wave frequencies \cite{liu2010}. The epitaxial technology
is less expensive, albeit relatively not as well developed. The
electron mobility is mainly determined by scattering on extended
defects and ionized impurities in heavily doped layers, while polar
scattering by longitudinal optical phonons (LO phonons) is important
at low and moderate electron densities \cite{look2006,liu2012}. The
electron mobility is not as high than in GaN but the value slightly
above 400 cm$^2/$(Vs) is achieved in epitaxial layers at room
temperature and low residual impurity densities
\cite{tsukazaki2006}.

The supplied electric power is dissipated mainly through electron
coupling with LO phonons, while the defect and ionized impurity
scattering is elastic and acoustic phonon scattering is
quasi-elastic. Because the high-field electron transport is
accompanied with intense LO-phonon emission by hot electrons, the
emitted non-equilibrium LO phonons accumulate, and the associated
phenomena are often referred to as hot-phonon effects. In general,
hot phonons play an important role in devices operated at high
electric fields such as microwave and high power field-effect
transistors because of pivotal role they play in heat dissipation
and device reliability \cite{matulionis2012}. The fastest decay of
hot phonons takes place in the vicinity of plasmon–-LO-phonon
resonance \cite{matulionis2010}. Essentially, the plasmon-assisted
ultrafast decay of hot phonons accompanied with very short LO-phonon
lifetimes causes fast electron energy relaxation
\cite{sermuksnis2015}, high electron drift
velocity\cite{liberis2014,Ardaravicius2011} and the best operation
of the device.

Accumulation of non-equilibrium (hot) longitudinal optical (LO)
phonons reduces the electron drift velocity but causes a weak effect
if their decay is fast \cite{Khurgin2007,dyson2010,liberis2014}.
Initial studies of high-field electron transport measurements in
epitaxial ZnO structures was reported in Ref.\cite{sasa2008}. The
dependence of electron drift velocity on applied electric field was
deduced at room temperature for nominally undoped MBE grown ZnO up
to 100 kV/cm \cite{sasa2011}. Although short enough voltages pulses
(300 ns) were used to heat the electrons, the self-heating effect
due to current caused relatively low drift velocity values
(7.6$\times10^6$ cm/s). No experimental studies of the velocity
dependence on the electron density was reported.

In this work the current--voltage dependence at different electron
densities was measured for (Ga,Sb) doped ZnO films at high electric
fields. Nearly-equilibrium acoustic phonon temperature was ensured
by the semi-automated few-nanosecond-pulsed voltage technique. The
electron drift velocities were deduced under assumption of uniform
electric field and constant electron density. Monte Carlo simulation
data obtained by taking hot phonons into account were used to
interpret experimental results.

\section{Samples}

The ZnO films were grown on a-plane sapphire substrates by
plasma-enhanced molecular beam epitaxy (PMBE). Plasma power of 400 W
was used for depositing of the films. The growth of the ZnO films
was monitored by in situ reflection high energy electron
diffraction. A 5-nm-thick low-temperature ZnO buffer layer was
deposited at  $300^{\circ}$ C for enhanced nucleation and 10
annealed at  $700^{\circ}$ C prior to the growth of doped ZnO layer.
The layer thickness was 150-–350 nm. Structural properties and
surface morphology were studied by X-ray diffraction, transmission
electron microscopy, and atomic force microscopy. The transmission
line model (TLM) patterns were processed with evaporated stacks of
Ti/Au (25~nm/30~nm ) acting as Ohmic contacts. The channel width $w$
was 250-300 $\mu$m and the length (inter-electrode distance) $L$ was
1.7, 2.9, 3.9, 5.8, 6.1, 6.9, 9.2, 9.9, 14.8, 16,8, 17.3 $\mu$m. The
Hall effect measurements were performed in the van der Pauw
configuration with soldered indium contacts. The Hall mobility $\mu$
values measured for the investigated channels ranged from
110$~\rm{cm}^2$/(V$\cdot$s) to 23$~\rm{cm}^2$/(V$\cdot$s). The 3DEG
density $n_{\rm 3D}$ assessed from Hall measurements ranged from
1.42$\times$10$^{17}$cm$^{-3}$ to 5.7$\times$10$^{20}$cm$^{-3}$. The
electron density, mobility values and the channel thickness $d$ are
listed in Table 1.
\begin{table}[h]
\caption{3DEG density, Hall mobility and channel thickness for the
ZnO films at room temperature.}
  \begin{tabular}[htbp]{ccccc}
    \hline
    \br
    Wafer&dopant& $n_{\rm 3D}$&$\mu$&$d$\\
    &&cm$^{-3}$&\,${\rm cm}^2$/V\,s&nm\\
    \hline
    \mr
    \#567 &Ga&$1.42\times 10^{17}$&106&330-340\\
    \#570 &Ga&$5.51\times 10^{17}$&73&350\\
    \#639 &Ga&$4.9\times 10^{18}$&23&300\\
    \#753 &Ga&$1\times 10^{19}$&57&200\\
    \#751 &Ga&$1.1\times 10^{17}$&66&200\\
    \#487 &Sb&$4.6\times 10^{19}$&110&170\\
    \#661 &Ga&$1.74\times 10^{19}$&25-30&150\\
    \#658 &Ga&$1.13\times 10^{20}$&48&350\\
    \#592 &Ga&$5.7\times 10^{20}$&40&380\\
    \hline
    \br
  \end{tabular}
  \label{onecolumntable}
\end{table}

The contact resistance $R_{\rm c}$ was estimated at low electric
fields from the dependence of the sample resistance on the channel
length.

\section{Methods}

The electron transport measurements at high electric fields were
carried out on two-electrode samples taken from the TLM structures.
The sample was placed into a gap of a micro-strip line. The
measurement procedure and the determination of current--voltage
$(I-V)$ dependence is described in
Refs.\cite{ardaravicius2009,ardaravicius2015}. The employment of
nanosecond pulses enabled us to minimize the complications caused by
channel self-heating due to current. The electron drift velocity was
estimated under assumptions that the electric field $E$ was uniform
and 2DEG density was independent on the electric field: $v_{\rm
dr}$= $I(E)/(en_{\rm 3D}w$), where $e$ is an elementary charge.

The channels suffered soft damage at high electric fields. The
damage was characterized by the change of the zero-field channel
resistance measured before and after the high-field experiment. We
present the results up to a damage level not exceeding 5~\%. The
soft damage turned into the channel breakdown at higher fields.

Ensemble Monte Carlo simulation was carried out for bulk wurtzite
ZnO crystals at 300 K lattice temperature. The electron motion and
scattering in the one-valley spherical parabolic conduction band was
considered. In the range of electric fields under investigation,
electron scattering into the upper valleys was expected to be
unimportant and the inter-valley transfer was neglected
\cite{furno2008}. The scattering mechanisms included in the
simulation were: acoustic phonon scattering, LO phonon scattering,
and ionized impurity scattering.

The hot-phonon effect was included in the Monte Carlo model for ZnO
in a manner described previously \cite{ramonas2005}. The LO-phonon
distribution was updated after each event of LO-phonon emission and
absorption, and after decay of the excess LO phonons. The decay into
modes of the unperturbed thermal bath was treated. The
time-dependent LO-phonon distribution was calculated under the
assumption of dispersion-less LO phonons. The decay of excess LO
phonons into other modes of the thermal bath was treated in the
LO-phonon lifetime approximation, the Ridley model was used
\cite{ridley1996}. The results of the Monte Carlo simulations are
used to calculate the mean electron energy, electron energy
autocorrelation function and electron drift velocity. The
simulations were carried out at a fixed background temperature of
acoustic phonons representing the thermal bath for the subsystem of
hot electrons and hot phonons. As mentioned, the channel background
temperature differs from the hot-electron temperature and the remote
heat sink temperature.

\section{Results and discussion}

Pulsed technique is used to investigate dependence of current on
voltage. Hot electrons and hot phonons are the main causes for
non-ohmic behaviour at high electric fields when the pulses of
voltage are sufficiently short. Because of strong electron-LO-phonon
coupling in ZnO, emission of LO phonons by hot electrons is the
preferred route for dissipation of the heat gained by the mobile
electrons from the applied electric field. The dissipation of the
electronic heat is partly compensated by re-absorption of the
emitted LO phonons. Thus the heat is shared by the hot electrons and
the hot phonons before it is converted into the heat modes able to
propagate across the electrically non-conductive layers towards the
remote heat sink. The bottleneck forms if the hot-phonon conversion
is slow and the associated hot-phonon lifetime is long. The
dependence of current on the pulse duration is associated with
increase in the background temperature. Nanosecond pulses of voltage
are used in order to minimize the thermal walkout as illustrated for
high electron density sample \#592 (Figure 1, see also
\cite{ardaravicius2003}).

%-------------------------------Fig.2
\begin{figure}[htb]
  \begin{center}
    \includegraphics[height=7.0cm]{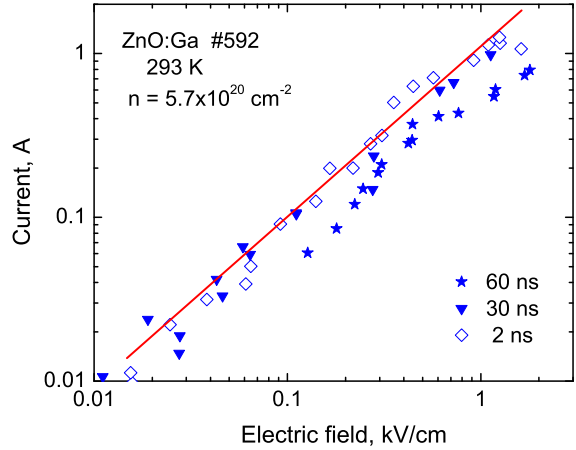}
    \end{center}
\caption{Current--field dependence for ZnO sample \#592 at electron
density of 5.7$\times$10$^{20}$ cm$^{-3}$. Voltage pulse duration 60
ns (stars), 30 ns (triangles), and 2 ns (diamonds).}\label{2pav}
\end{figure}
The experimental data for sample \#567 are illustrated in Figure 2
(closed squares, triangles).
%-------------------------------Fig.3
\begin{figure}[htb]
  \begin{center}
    \includegraphics[height=7.0cm]{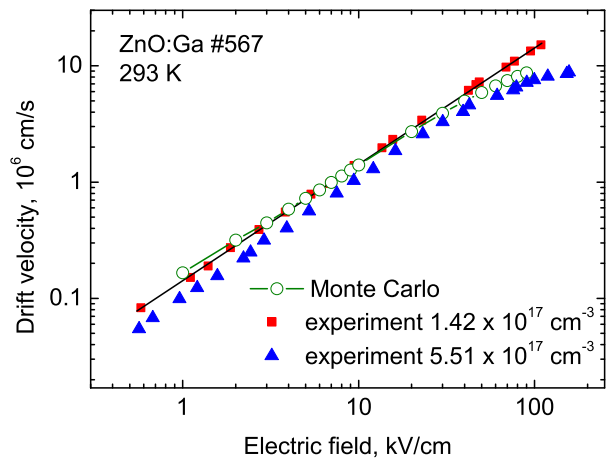}
    \end{center}
\caption{Dependence of drift velocity on applied electric field for
ZnO sample \#567 (closed squares, triangles). Voltage pulse duration
is 2 ns. Monte Carlo simulation (open circles) for hot-phonon
lifetime of 1 ps.} \label{3pav}
\end{figure}
The highest electric field of 150 kV/cm is reached for electron
density of 5.51$\times$10$^{17}$ cm$^{-3}$ (blue triangles). The Ohm
law (black line) approximately holds at electric fields up to 100
kV/cm at the electron density of 1.42$\times$10$^{17}$ cm$^{-3}$
(red squares). The drift velocity is $\sim$1.5$\times$10$^{7}$ cm/s
at 104 kV/cm. This value for the drift velocity exceeds the recently
published experimental result of $\sim$7.6$\times$10$^{6}$ cm/s at
around 100 kV/cm at the electron density of
$\sim$2.4$\times$10$^{16}$ cm$^{-3}$ measured in nominally undoped
ZnO structures grown by MBE \cite{sasa2008,sasa2011}.

Sample \#487 with a higher electron density of  4.6$\times$10$^{19}$
cm$^{-3}$ demonstrates  lower values for the drift velocity (Fig. 3,
red squares).  The velocity is close to $\sim$3$\times$10$^{5}$ cm/s
at 6 kV/cm for sample \#487 (Fig. 3, squares) in contrast to a
higher value of $\sim$8.5$\times$10$^{5}$ cm/s at the same electric
field for sample \#567 (Fig. 2, squares). The application of 2 ns
voltage pulses enabled us to measure the drift velocity at electric
fields up to 150 kV/cm (Fig. 2, triangles).
%-------------------------------Fig.3
\begin{figure}[htb]
  \begin{center}
    \includegraphics[height=7.0cm]{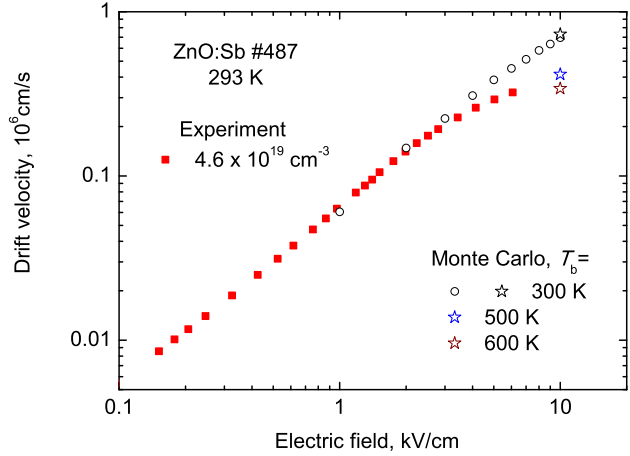}
    \end{center}
\caption{Dependence of drift velocity on applied electric field for
ZnO sample \#487 (closed squares). Voltage pulse duration is 2 ns.
Monte Carlo simulation: at room temperature for hot-phonon lifetime
of 1 ps (open circles); at elevated background temperatures: 300 K
(open circles and open black star), 500 K (open blue star), and 600
K (open wine star).} \label{3pav}
\end{figure}
The experimental electron drift velocity increases linearly with the
applied electric field at electron density of 1.42$\times$10$^{17}$
cm$^{-3}$ (Fig. 2). This behavior fortuitously indicates a weak
impact of the hot phonons on the electron transport. On the other
hand, an essential deviation from the Ohm law is observed at
electron density of 4.6$\times$10$^{19}$ cm$^{-3}$ (Fig. 3). The
experimental results are compared with those of Monte Carlo
simulations.

The impurity scattering is taken into account in order to fit the
experimental data at low electric field. The density of impurities
is artificially increased over the density of electrons-high
impurity/electron density ratio is used, the compensation ratio is
from 10 to 13. This is an important argument for strong elastic
electron scattering in the investigated samples. Since we aim at
investigation of hot-electron energy relaxation, we feel safe to
mimic scattering on impurities, defects and lattice imperfections by
ionized impurity scattering: the scattering on both charged and
neutral lattice imperfections is usually assumed to be elastic. The
results of Monte Carlo simulation are illustrated in Figures 2 and 3
under a realistic assumption that the hot-phonon lifetime is equals
1 ps (open circles and stars). The comparison with the experimental
results suggests that the lifetime is below 1 ps for sample \#567
(squares).

The short voltage pulses help to reduce  the thermal walkout effect
unless the electron density is high and the electric field is
strong. Figure 3 presents the results at high electron density
(4.6$\times$10$^{19}$ cm$^{-3}$) in the electric field range below 6
kV/cm. The thermal walkout turns into the electrothermal breakdown
at stronger electric fields even when the voltage pulses of 2 ns are
applied. Stars in Fig. 3 illustrate the simulated hot-electron drift
velocity at different background temperatures. The drift velocity
decreases as the background temperature increases. Therefore, the
experimental deviation from the Ohm law cannot be interpreted in
terms hot-electron energy relaxation. An interpretation in terms of
elevated background temperature seems quite probable (cf. closed
square and wine star in Figure 3). The alternative interpretation of
the experimental data in terms of hot-phonon lifetime requires
unrealistically long lifetimes-significantly longer than 1 ps-if the
thermal walkout is ignored.

The enhanced resonance decay of hot phonons has a direct consequence
on various phenomena where the excess occupancy of hot-phonon modes
plays an important role
\cite{matulionis2010,Ardaravicius2011,matulionis2012,matulionis2013}.
In particular, the electron scattering by the non-equilibrium excess
(hot) phonons is weaker if the occupancy of the involved hot-phonon
modes is lower. This can be demonstrated by measuring the
hot-electron drift velocity in various samples at the chosen value
of applied electric field.
%-------------------------------Fig.5
\begin{figure}[htb]
  \begin{center}
    \includegraphics[height=7.0cm]{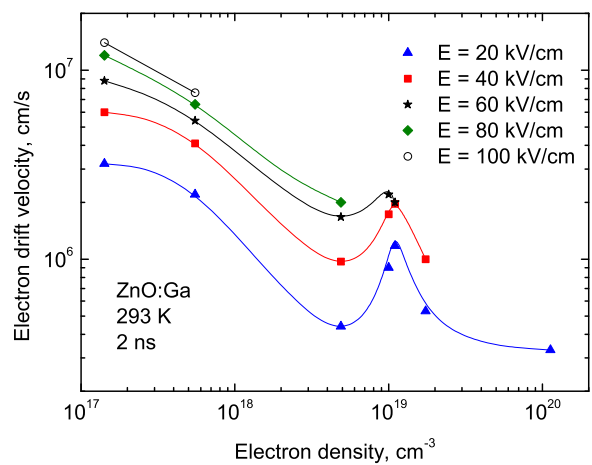}
    \end{center}
\caption{Experimental dependence of hot-electron drift velocity on
electron density for doped ZnO. Lines guide the eye. Data at high
fields for high-density channels are missing because of thermal
runaway. Voltage pulse duration 2 ns.} \label{5pav}
\end{figure}
Typically, the effect is masked by the Joule effect: the increase in
background temperature causes a stronger scattering of electron by
acoustic phonons. The Joule effect is important at high electric
field in channels with high electron density. Application of short
pulses of voltage allows one to reduce the thermal walkout effect
dramatically. Let us plot the density dependence of the drift
velocity  (Fig. 4). Typically, the drift velocity decreases if the
doping increases. This happens in the electron density range from
1$\times$10$^{17}$ cm$^{-3}$  to 5$\times$10$^{18}$ cm$^{-3}$  for
doped ZnO (Fig. 4, symbols). The velocity stops decreasing at around
5$\times$10$^{18}$ cm$^{-3}$, and the local maximum forms at higher
densities at intermediate electric fields in the range from 20 kV/cm
to 60 kV/cm (triangles, squares, stars). The thermal runaway makes
it impossible to resolve the maximum at electric fields of 80 kV/cm
and 100 kV/cm (diamonds, circles). The maximum is located at above
10$^{19}$ cm$^{-3}$. Its position (Fig. 4) correlates with that of
the resonance decay of hot phonons \cite{sermuksnis2015}. Thus, the
observed fast decay of hot phonons supports high values of
hot-electron drift velocity unless the thermal runaway comes into
play.

\section{Conclusions}

The nanosecond pulsed measurements yielded hot-electron drift
velocity in agreement with those of the Monte Carlo simulation when
the density of the ionized impurities is allowed to exceed that of
the electrons by 10-–13 times in bulk ZnO. The Ohm law approximately
held in the electric field range up to 100 kV/cm at an electron
density of 1.42$\times$10$^{17}$ cm$^{-3}$ but the range shrank at
higher electron densities. The LO-phonon--plasmon resonance
supported increase in the electron drift velocity is resolved at
($\sim$10$^{19}$ cm$^{-3}$). The non-ohmic behavior at higher
densities ($>$10$^{19}$ cm$^{-3}$) is caused by increase in the
background temperature represented by acoustic phonons. A drift
velocity of $\sim$1.5$\times$10$^{7}$ cm/s was measured at 104 kV/cm
field. A lot of effort still to be done to design ZnO-based
field-effect transistors with improved electron transport along a
channel.

\section*{Acknowledgement}
This research is funded by the Research Council of Lithuania (grant
No. APP-5/2016).

\newcommand{\newblock}{}


\begin{thebibliography}{99}

\bibitem{morkoc2009}
Morko\c{c} H 2009, {\it Zinc Oxide: Fundamentals, Materials and
Device Technology} (Weinheim: Wiley-VCH)

\bibitem{liu2010}
Liu H, Avrutin V, Izyumskaya N, \"{O}zg\"{u}r \"{U}, Morko\c{c} H
2010 {\it Supperlattices Microstruct.} {\bf 48} 458

\bibitem{look2006}
Look D C, Leedy K D, Tomich D H, and Bayraktaroglu B 2006 {\it Appl.
Phys. Lett.} {\bf 96} 062102

\bibitem{liu2012}
Liu Y, Avrutin V, Izyumskaya N, \"{O}zg\"{u}r \"{U}, Yankovich A B,
Kvit A V, Voyles P M, and Morko\c{c} H 2012 {\it J. Appl. Phys} {\bf
111} 103713

\bibitem{tsukazaki2006}
Tsukazaki A, Ohtomo A, and Kawasaki M 2006 {\it Appl. Phys. Lett.}
{\bf 88} 152106

\bibitem{matulionis2012}
Matulionis A, Liberis J, \v{S}ermuk\v{s}nis E, Ardaravi\v{c}ius L,
\v{S}imukovi\v{c} A, Kayis C, Zhu C Y, Ferreyra R, Avrutin V,
\"{O}zg\"{u}r \"{U}, Morko\c{c} H  2012 {\it Microelectron. Reliab.}
{\bf 52} 2149
%----------------------------------1
\bibitem{matulionis2010}
Matulionis A, Liberis J, Matulionien\.{e} I, Ramonas M, and 2010
{\it Proc. IEEE} {\bf 98} 1118

\bibitem{sermuksnis2015}
\v{S}ermuk\v{s}nis E, Liberis J, Ramonas M, Matulionis A, Toporkov
M, Liu H Y, Avrutin V, \"{O}zg\"{u}r \"{U}, and Morko\c{c} H 2015
{\it J. Appl. Phys.} {\bf 117} 065704

\bibitem{liberis2014}
Liberis J, Ramonas M, \v{S}ermuk\v{s}nis E, Sakalas P, Szabo N,
Schuster M, Wachowiak A, and Matulionis A 2014 {\it Semicond. Sci.
Technol.} {\bf 29} 045018

%--------------------------------- 8
\bibitem{Ardaravicius2011}
Ardaravi\v{c}ius L, Liberis J, Kiprijanovi\v{c} O, Matulionis A, Wu
M, and Morko\c{c} H 2011 {\it Phys. Status Solidi RRL} {\bf 5} 65

%--------------------------------- 10

\bibitem{Khurgin2007}
Khurgin J, Ding Y J, and Jena D 2007 {\it Appl. Phys. Lett.} {\bf
91} 252104

\bibitem{dyson2010}
Dyson A and Ridley B K 2010 {\it J. Appl. Phys.} {\bf 108} 104504

\bibitem{sasa2008}
Sasa S, Hayafuji T, Kawasaki M, Nakashima A, Koike K, Yano M, and
Inoue M 2008  {\it Phys. Status Solidi C} {\bf5} 115

\bibitem{sasa2011}
Sasa S, Maitani T, Furuya Y, Amano T, Koike K, Yano M, and Inoue M
2011 {\it Phys. Status A} {\bf 208} 449

\bibitem{ardaravicius2009}
Ardaravi\v{c}ius L, Ramonas M, Liberis J, Kiprijanovi\v{c} O,
Matulionis A, Xie J, Wu M, Leach J H, and Morko\c{c} H 2009 {\it J.
Appl. Phys.} {\bf 106} 073708

\bibitem{ardaravicius2015} Ardaravi\v{c}ius L, Kiprijanovi\v{c} O,
Liberis J, ~Matulionis A, \v{S}ermuk\v{s}nis E, Ferreyra R A,
Avrutin V, \"{O}zg\"{u}r \"{U}, and Morko\c{c} H 2015 {\it Semicond.
Sci. Technol.} {\bf 30} 105016

\bibitem{furno2008}  Furno E, Bertazzi F, Goano M, Ghione G, and
Bellotti E 2008 {\it Solid-State Electron.} {\bf 52} 1796

\bibitem{ramonas2005} Ramonas M,  Matulionis A, Liberis J, Eastman L F, Chen X,
and Sun Y J 2005 {\it Phys. Rev. B} {\bf 71} 075324

\bibitem{ridley1996}
Ridley B K 1996 {\it J. Phys.: Condens. Matter} {\bf 8} L511

\bibitem{ardaravicius2003} Ardaravi\v{c}ius L,~Matulionis
A,~Liberis J, Kiprijanovic O., Ramonas M, Eastman L~F, Shealy J. R,
and Vertiatchikh A 2003 {\it Appl. Phys. Lett.} {\bf 83} 4038

\bibitem{matulionis2013}
Matulionis A 2013 {\it Semicond. Sci. Technol.} {\bf 28} 074007

%\bibitem{masetti1983}
%Masetti G, Severi M, and Solmi S 1983 {\it IEEE Trans. Electron
%Devices} {\bf 30} 764

%\bibitem{reggiani2002}
%Reggiani S, Valdinoci M, Colalongo L, Rudan M, Baccarani G, Stricker
%A D, Illien F, Felber N, Fichtner W, and Zullino L 2002 {\it IEEE
%Trans. Electron Devices} {\bf 49} 490

%\bibitem{ridley2013}
%Ridley B~K 2013 {\it Quantum Processes in Semiconductors} (Oxford:
%Oxford University Press) 296




\end{thebibliography}
\end{document}